\documentclass[preprintnumbers,superscriptaddress,showkeys,showpacs,byrevtex]{revtex4}
\usepackage{amsmath,amsfonts,amssymb,amscd,amsxtra,amsthm}
\usepackage{graphicx}
\usepackage{epstopdf}
\usepackage{bm}
\begin{document}
\preprint{PKNU-NuHaTh-2017-03}
\title{Photo- and electroproduction of $\Lambda(1405)$ via $\gamma^{(*)}p\to K^+\pi^+\Sigma^-$}
\author{Seung-il Nam}
\email[E-mail: ]{sinam@pknu.ac.kr}
\affiliation{Department of Physics, Pukyong National University (PKNU), Busan 608-737, Republic of Korea}
\affiliation{Asia Pacific Center for Theoretical Physics (APCTP), Pohang 790-784, Republic of Korea}
\date{\today}
\begin{abstract}
In this work, we perform a \textit{toy-model} analysis for the unpolarized photo- and electroproduction of $\Lambda(1405)\equiv\Lambda^*$ via $\gamma^{(*)}p\to K^+\pi^+\Sigma^-$ by employing the effective Lagrangian approach at the tree level only. We consider that  $\Lambda^*$ consists of the high-mass ($H$) and low-mass ($L$) poles as suggested by the chiral unitary model (ChUM). We determine all the model parameters including coupling constants and cutoff parameters for the phenomenological strong form factors by using the theoretical information from ChUM and available experimental data. The electromagnetic (EM) form factors for the two poles are parameterized appropriately being similar to that for the neutron by employing the ChUM estimates for the EM root-mean-squared (rms) radii for the two poles. We observe from the numerical calculations that, for the photoproduction of $\Lambda^*$, the interference between the two poles turns out to be destructive, resulting in the single-peak line at $M_{\pi^+\Sigma^-}\approx1405$ MeV in the invariant-mass distribution. On the contrary, there appear two peaks in the distribution for the electroproduction as observed in the CLAS/Jlab electroproduction data, due to the constructive interference, which is mainly caused by the Dirac form factors for the two poles. From this observation, we conclude that, in order to explain the photo- and electroproduction data simultaneously, i.e., \textit{single} and \textit{double} peaks, respectively, in the invariant-mass distribution, the interference patterns for the two poles should follow those suggested by ChUM. In turn, it is a strong theoretical and experimental supports for the two-pole structure scenario for $\Lambda^*$.
\end{abstract}
\maketitle
\section{Introduction}
The tetraquarks, pentaquarks, and meson-baryon molecular states, i.e., the \textit{exotics} are one of the most interesting objects to be studied in the hadron spectra governed by the nonperturbative quantum chromodynamics (QCD). Recent progresses on the researches for those exotics can lead us to the new understanding of nonperturbative QCD. The four-quark meson, i.e., tetraquark state, has been reported by the Belle Collaboration and BESIII Collaboration~\cite{Choi:2003ue,Abe:2007jna,Choi:2007wga,Belle:2011aa,Ablikim:2013mio,Liu:2013dau,Ablikim:2013wzq}. Recently, the LHCb Collaboration measured  signals for the heavy pentaquark state $P_c^+$ as well~\cite{Aaij:2015tga}. The meson-baryon molecular state for $\Lambda(1405)\equiv\Lambda^*$, rather than a simple three-quark ($uds$) one, was suggested  from the unitarized chiral-dynamics models~\cite{Dalitz:1959dn,Nacher:1998mi,Jido:2002zk,Jido:2003cb,Nam:2003ch,Lutz:2004sg,Magas:2005vu,Hyodo:2011ur,Mai:2014xna}, and supported by the lattice-QCD (LQCD) simulation by investigating the strange form factor of $\Lambda^*$~\cite{Nemoto:2003ft}. In addition to the studies of structure for the $\Lambda(1405)$, the its production processes were also  scrutinized extensively as well in Refs.~\cite{Nam:2008jy,Nakamura:2013boa,Nam:2015yoa,Wang:2016dtb}. In order to study the internal structure of $\Lambda^*$, the constituent-counting rule, which provides different angular behavior in the high-energy region depending on the number of quarks inside hadrons, was employed In Refs.~\cite{Kawamura:2013iia,Chang:2015ioc}.

Note that the line-shape analysis for the $\Lambda^*$ invariant-mass distribution is also an important and interesting research subject, since the theories and experiments have not been consistent to each other and it depends much on the isospin channels of $\pi\Sigma$~\cite{Ahn:2003mv,Niiyama:2009zza,Moriya:2013hwg,Moriya:2013eb}. Nonetheless, although the quantitative and detailed line shapes are different in many ways, qualitatively, there appears single-peak line shape in common for the photoproduction of $\Lambda^*$ as in the ChUM calculations~\cite{Jido:2002zk,Nam:2003ch,Hyodo:2011ur}. In Ref.~\cite{Lu:2013nza}, however, the CALS/Jlab collaboration reported the electroproduction of $\Lambda^*$ showing very distinctive line shape, i.e., the double-peak one, and  concluded that this nontrivial line shape is an evidence for the two-pole structure for $\Lambda^*$, due to the different excitation of each pole depending on the photon virtuality $Q^2$. Hence, in the present work, we would like to study this obvious differences between the photo- and electroproduction by using a simplified \textit{toy-model}, focusing on the interference patterns between the two poles. For this purpose, we want to work with the Dalitz process $\gamma^{(*)}p\to K^+\pi^+\Sigma^-$.

As mentioned, we simplify the production process of $\Lambda^*$ as follows: 1) As for the possible intermediate hyperons, which decay into $\pi\Sigma$, we only consider the $\Lambda^*$ resonance, assumed to be composed of the two poles in order to reduce theoretical uncertainties. Although the $\Sigma^*(1385)$ resonance may play an important role and interferes with the low-mass pole, as suggested in Ref.~\cite{Jido:2009jf}, we ignore it rather safely, because its contribution turns out to be almost negligible from the CLAS/Jlab electroproduction data~\cite{Lu:2013nza}. 2) In the present production process, we can ignore the interferences between the intermediate $K^*$ meson, which decays into $\pi^+ K^+$, and the $\Lambda^*$ baryon on the Dalitz plot. This exclusion of $K^*$ makes the present analysis much simple and obvious. 3) The far-off-shell diagrams, in which the incident photon couples to the final-state particles ($K^+$, $\pi^+$, and $\Sigma^-$), provide very small contributions to the total cross section. We verified that those contributions are below $10\%$ to the total. Hence, we exclude those diagrams for the present work for brevity, while the Ward-Takahashi identity (WTI) remains satisfied. 4) From our previous works~\cite{Kim:2017nxg}, it turned out that the nucleon resonances become almost negligible in describing the invariant-mass spectra, if we are away from the reaction threshold, i.e., $\sqrt{s}\ge 2.4$ GeV for instance. Thus, focusing on the higher $\sqrt{s}$ region, we do not include the nucleon resonances, which can cause many theoretical uncertainties. 5) The $K^*$ exchange in the $t$ channel is also omitted, because of its small coupling constants to the $\Lambda^*$~\cite{Khemchandani:2011mf} as verified in our previous work~\cite{Khemchandani:2011mf}. Although some important ingredients could be missing from these  simplifications, one can clearly identify and separate the crucial origin for the difference between the photo- and electroproduction of $\Lambda^*$ without loss of generality.

In this setup, we employ the effective Lagrangian approach at the tree level for computing the invariant-mass distribution for $\gamma^{(*)}p\to K^+\pi^+\Sigma^-$ as a function of $M_{\pi^+\Sigma^-}$, since we are interested in the single- and double-peak line shapes observed in the photo- and electroproduction, respectively. $\Lambda^*$ is considered as a composite of the high-mass ($H$) and low-mass ($L$) poles as suggested by ChUM. All the model parameters including coupling constants and cutoff parameters for the phenomenological strong form factors are determined by using the theoretical information from ChUM and available experimental data~\cite{Jido:2002zk,Nam:2003ch,Hyodo:2011ur,Khemchandani:2011mf,Olive:2016xmw}. We try to parameterize the electromagnetic (EM) form factors for the two poles being similar to that for the neutron by employing the ChUM estimates for the EM root-mean-squared (rms) radii for the two poles~\cite{Sekihara:2008qk}. 

It turns out that, for the photoproduction of $\Lambda^*$, the interference between the two poles becomes destructive, resulting in the single-peak line shape at $M_{\pi^+\Sigma^-}\approx1405$ MeV in the invariant-mass distribution as expected. On the contrary, there appear double-peak one in the distribution for the electroproduction as observed in the CLAS/Jlab electroproduction data, due to the constructive interference, which is mainly caused by the \textit{Dirac} form factors for the two poles. In detail, the relative phase factor between the high-mass and low-mass pole amplitudes becomes almost opposite for the photoproduction, in comparison to the electroproduction. Consequently, in order to explain the photo- and electroproduction data for the invariant-mass distributions simultaneously, the interference patterns for the two poles should follow those suggested by ChUM. Again, this observation can be a strong theoretical and experimental support for the exotic two-pole structure for $\Lambda^*$.

The present work is organized as follows: In Section II, we will make a brief explanation for the theoretical framework and define the effective Lagrangians, phenomenological strong and EM form factors, and so on. The numerical results and relevant discussions are given in Section III. Section IV is devoted to summary and future perspective. 
\section{Theoretical Framework}
\begin{figure}[h]
\includegraphics[width=14cm]{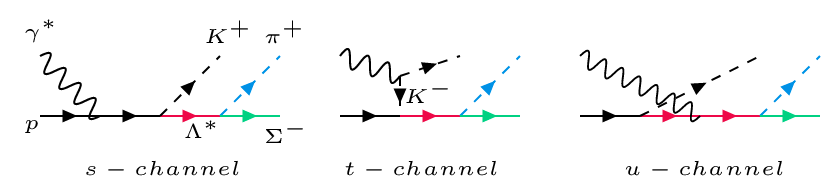}
\caption{(Color online) Relevant Feynman diagrams for $\gamma^*(k_1)\,p(k_2)\to K^+(k_3)\,\pi^+(k_4)\,\Sigma^-(k_5)$ producing $\Lambda^*$ with the four momenta $k_i$.}       
\label{FIG0}
\end{figure}

In this Section, we make a brief explanation for our theoretical framework. First, we depict the relevant Feynman diagrams for the present production process in Fig.~\ref{FIG0}. We define the four momenta for the particles involved by $k_{1,2,3,4,5}$ for the incident photon, the target proton, the outgoing kaon, the outgoing pion, and the sigma baryon, respectively. As mentioned previously, we only focus on these three topologically-independent diagrams, in which the incident photon couples to the proton, the kaon and the $\Lambda^*$, because these diagrams dominate the process. The effective Lagrangians for the EM and strong interaction vertices in the Feynman diagrams are defined by~\cite{Kim:2017nxg} 
\begin{eqnarray}
\mathcal L_{\gamma \Phi \Phi} &=& 
-ie_\Phi [  \Phi^\dagger (\partial_\mu \Phi) - (\partial_\mu \Phi^\dagger) \Phi] A^\mu ,   
 \cr  
\mathcal L_{\gamma BB} &=&
- \bar B \left[ e_B \gamma_\mu - \frac{e\kappa_B}{2M_N}
\sigma_{\mu\nu}\partial^\nu \right] A^\mu B,      
\cr
\mathcal L_{\Phi BB} &=&ig_{\Phi BB'}\bar{B}\Phi B',                     
\label{eq:BornLag1}
\end{eqnarray}
where $\Phi$, $B$, and $A$ denote the pseudoscalar meson, the baryon, and the photon fields, respectively. $e_B$ and $e_\Phi$ stand for the electric charge for the baryon and the meson, whereas $e$ indicates the unit electric charge. The anomalous magnetic moment for the baryon is given by $\kappa_B$. Note that we do not have $\gamma_5$ in the strong interaction Lagrangian, because all the baryon-baryon coupling in the present work is that of the positive and negative baryons, i.e., $p(1/2^+)$-$\Lambda^*(1/2^-)$ or $\Sigma(1/2+)$-$\Lambda^*(1/2^-)$. Note that, as for the $\Phi BB'$ interaction vertex, we employ the pseudo-scalar (PS) meson-baryon coupling scheme, although the pseudo-vector (PV) one is also alternatively possible. Taking into account the discussion given in Ref.~\cite{Nam:2003uf}, the difference between the two schemes is in principle proportional to
\begin{equation}
\label{eq:DIF}
i\mathcal{M}_\mathrm{PV}-i\mathcal{M}_\mathrm{PS}\equiv i\Delta\mathcal{M}\propto e\left(\frac{\kappa_B}{2M_B}+\frac{\kappa_{B'}}{2M_{B'}}\right)\rlap{/}{\epsilon}\rlap{/}{k}_1,
\end{equation}
and this \textit{magnetic} contribution is highly suppressed in the low-energy region  with a phenomenological form factor scheme to reproduce data~\cite{Nam:2003uf}. Hence, we will perform all the numerical calculations with the PS scheme in the present work.

Relevant numerical inputs for the particles are given in Table~\ref{TAB0}. All the coupling constant for the $H$ and $L$ poles are taken from the theoretical estimates from ChUM~\cite{Jido:2002yz,Sekihara:2008qk}, and the same for the anomalous magnetic moments $\kappa_{H,L}$~\cite{Jido:2003cb}. The pole positions for $H$ and $L$ are slightly different from those suggested by ChUM~\cite{Jido:2003cb}, because we employ the values extracted from the CLAS/Jlab electroproduction data~\cite{Lu:2013nza}. Moreover, the full-decay widths are modified to reproduce the CLAS/Jlab photoproduction data~\cite{Moriya:2013hwg}, in comparison to those from ChUM~\cite{Jido:2003cb}. Here, it is worth mentioning that the coupling constants for $H$ and $L$ are given in complex numbers~\cite{Jido:2003cb}. 
\begin{table}[b]
\begin{tabular}{|c|c|c|c|c|c|}
\hline
$B$&Mass&Width&$g_{KNB}$&$g_{\pi\Sigma B}$&$\kappa_B$\\
\hline
$p$&$938.272$ MeV&$\sim0$ MeV&$-$&$-$&$2.79$\\
\hline
$L$&$1368\,$ MeV&$100\,$ MeV&$1.2+1.7i$&$-2.5-1.5i$&$0.30$\\
\hline
$H$&$1423\,$ MeV&$50\,$ MeV&$-2.5+0.94i$&$0.42-1.4i$&$0.41$\\
\hline
\end{tabular}
\caption{Relevant numerical inputs for the high-mass $\Lambda^*$ ($H$) and the low-mass $\Lambda^*$ ($L$) from the chiral unitary model~\cite{Jido:2003cb,Sekihara:2008qk,Jido:2002yz}.}
\label{TAB0}
\end{table}

Employing the effective Lagrangians defined in Ref.~\cite{Kim:2017nxg}, it is straightforward to derive the following invariant amplitudes for the present Dalitz process $\gamma^{*}p\to K^+\pi^+\Sigma^-$:
\begin{equation}
\label{eq:INVTOT}
i\mathcal{M}_\mathrm{total}=\sum_{\Lambda^*=H,L}\sum_{x}i\mathcal{M}^{\Lambda^*}_x,\,\,\,\,
i\mathcal{M}^{\Lambda^*}_x=ieg_{KN\Lambda^*}g_{\pi\Sigma\Lambda^*}
\frac{\bar{u}(k_5)[\rlap{/}{q}_{4+5}+M^2_{\Lambda^*}]
\Gamma^{\Lambda^*}_xu(k_2)}
{M^2_{\pi^+\Sigma^-}-M^2_{\Lambda^*}-i\Gamma_{\Lambda^*}M_{\Lambda^*}},
\end{equation}
where $x=(s,t,u)$. The generic interaction structures for each channel are defined as follows:
\begin{eqnarray}
\label{eq:EMV1}
\Gamma^{\Lambda^*}_s&=&
\frac{\rlap{/}{k}_1+\rlap{/}{k}_2+M_N}{s-M^2_N}\left[\rlap{/}{\epsilon}
+(F^p_1-1)\left[\rlap{/}{\epsilon}+\frac{(\epsilon\cdot k_1)\rlap{/}{k}_1}{Q^2+\delta}\right]
-\frac{\kappa_pF^p_2}{2M_N}(\rlap{/}{\epsilon}\rlap{/}{k}_1-\epsilon\cdot k_1)\right]F_c,
\cr
\Gamma^{\Lambda^*}_t&=&\frac{1}{t-M^2_K}
\left[\epsilon\cdot (2k_3-k_1)+(F_K-1)\left[\epsilon\cdot (2k_3-k_1)+\frac{(\epsilon\cdot k_1)[k_1\cdot (2k_3-k_1)]}{Q^2+\delta}\right]\right]F_c,
\cr
\Gamma^{\Lambda^*}_u&=&F_u
\left[F^{\Lambda^*}_1\left[\rlap{/}{\epsilon}
+\frac{(\epsilon\cdot k_1)\rlap{/}{k}_1}{Q^2+\delta}\right]
-\frac{\kappa_{\Lambda^*}F^{\Lambda^*}_2}{2M_N}(\rlap{/}{\epsilon}\rlap{/}{k}_1-\epsilon\cdot k_1)\right]
\frac{\rlap{/}{k}_2-\rlap{/}{k}_3+M_{\Lambda^*}}{u-M^2_{\Lambda^*}},
\end{eqnarray}
where the photon virtuality $Q^2\equiv -k^2_1$ and $\delta$ stands for a small value $\delta\in\textbf{R}\sim0$ to avoid singularities for $Q^2=0$. We assign $q_{i\pm j}\equiv (k_i\pm k_j)$ and the Mandelstam variables $s=(k_1+k_2)^2$ , $t=(k_1-k_3)^2$, and $u=(k_2-k_3)^2$ for convenience.  The antisymmetric tensor is given by
\begin{equation}
\label{eq:SIGMA}
\sigma_{\mu\nu}\epsilon^\mu k^\nu_1
=\frac{i}{2}(\rlap{/}{\epsilon}\rlap{/}{k}_1-\rlap{/}{k}_1\rlap{/}{\epsilon})
=i(\rlap{/}{\epsilon}\rlap{/}{k}_1-\epsilon\cdot k_1).
\end{equation}
Note that we wrote the above amplitudes with the EM and strong form factors explicitly. The EM form factors are the Dirac $(F_1)$ and Pauli $(F_2)$ ones for the meson and baryons, where as the strong ones are $F_{c,u}$. We will define those form factors in detail in Section III. In general, it is an important issue to consider the Ward-Takahashi identity (WTI) in reaction processes with the photons. It is easy to check that the total amplitude satisfies the WTI for the photo- and electroproduction using Eqs.~(\ref{eq:INVTOT}) and (\ref{eq:EMV1}). In order to verify the WTI for the present calculation, we replace $\epsilon$ into $k_1$ for the EM vertices in Eq.~(\ref{eq:EMV1}), resulting in
\begin{eqnarray}
\label{eq:EMV2}
\Gamma^{\Lambda^*}_{s,\epsilon\to k_1}&=&
\frac{-Q^2+2k_1\cdot k_2}{-Q^2+2k_1\cdot k_2}\left[1
+(F^p_1-1)\left[1-\frac{Q^2}{Q^2+\delta}\right]\right]F_c,
\cr
\Gamma^{\Lambda^*}_{t,\epsilon\to k_1}&=&\frac{Q^2+2k_1\cdot k_3}{-Q^2-2k_1\cdot k_3}
\left[1+(F_K-1)\left[1-\frac{Q^2}{Q^2+\delta}\right]\right]F_c
\cr
\Gamma^{\Lambda^*}_{u,\epsilon\to k_1}&=&
F_uF^{\Lambda^*}_1\left[1
-\frac{Q^2}{Q^2+\delta}\right]\rlap{/}{k}_1
\frac{\rlap{/}{k}_2-\rlap{/}{k}_3+M_{\Lambda^*}}{u-M^2_{\Lambda^*}}.
\end{eqnarray}
The EM form factors follow the normalization conditions as $F_1^p(0)=F_K(0)=1$ and $F^{\Lambda^*}_1(0)=0$ for $Q^2=0$. Thus, it is easy to check that $(\Gamma^{\Lambda^*}_{s,\epsilon\to k_1}+\Gamma^{\Lambda^*}_{t,\epsilon\to k_1}+\Gamma^{\Lambda^*}_{u,\epsilon\to k_1})=(F_c-F_c+0\times F_u)=0$ always for any $Q^2$ value, i.e., $Q^2=0$ (photoproduction) or $Q^2\ne0$ (electroproduction), satisfying the WTI.
 
With the frame-independent transverse-polarization parameter $\varepsilon$, which measures the strength of the transverse polarization in the virtual photon, the photon-polarization vectors, two transverse $(x,y)$ and one longitudinal $(z)$, are given by~\cite{Nam:2013nfa}
\begin{equation}
\label{eq:POL}
\varepsilon_x=\left(0,\sqrt{1+\varepsilon},0,0\right),\,\,\,\,
\varepsilon_y=\left(0,0,\sqrt{1-\varepsilon},0\right),\,\,\,\,
\varepsilon_z=\frac{\sqrt{2\varepsilon}}{\sqrt{|Q^2|}}\left(k,0,0,E_1\right).
\end{equation}
Note that $\varepsilon$ varies from $0.3$ to $0.7$ in the CLAS/Jlab experiment~\cite{Lu:2013nza}. In the present work, we fix it with $\varepsilon=0.5$ as a trial. The incident-particle energies $E_{1,2}$ in general expressions for $Q^2\ne0$ read
\begin{equation}
\label{eq:EGN}
E_1=\frac{s-Q^2-M^2_2}{\sqrt{s}},\,\,\,\,
E_2=\frac{s+Q^2+M^2_2}{\sqrt{s}}.
\end{equation}

Taking into account all these ingredients discussed so far, the double differential cross section for $\gamma^{(*)}p\to K^+\pi^+\Sigma^-$ in the center-of-mass (cm) frame reads
\begin{eqnarray}
\label{eq:DDCS}
\frac{d^2\sigma_{\gamma^{(*)}p\to K^+\pi^+\Sigma^-}}
{dM_{K^+\pi^+}dM_{\pi^+\Sigma^-}}
&=&\frac{1}{4|\vec{k}_{\gamma^{(*)}}|\left|E_{\gamma^{(*)}}-E_N\right|}
\frac{1}{128\pi^4s}
\int M_{K^+\pi^+}M_{\pi^+\Sigma^-}\,d\cos\theta_{K^+}\,
d\phi_{\Sigma^-}\,\frac{1}{2n}\sum|\mathcal{M}_{\gamma^{(*)}p\to K^+\pi^+\Sigma^-}|^2,
\end{eqnarray}
where $\theta_K$ denotes the scattering angle for the outgoing kaon in the cm frame. The summation runs over the target spin and photon polarization.  Here, we choose the number of the photon polarizations $n=(2,3)$ for the photo- and electroproduction, respectively. The factor $1/(2n)$ indicates the average for the initial spin-polarization states.

\section{Strong and electromagnetic form factors}
It has been well-known that hadrons are spatially extended objects so that phenomenological form factors play important roles. In the present work, we consider the strong and EM form factors for the EM vertices, to which the incident photon couples, and the strong vertices for $KN\Lambda^*$ as shown in Fig.~\ref{FIG0}. For brevity, we do not consider the strong form factors for the $\pi\Sigma\Lambda^*$ vertices, due to their far off-shell nature, resulting in that the form factors are almost unity.

As for the strong form factor, we employ that used in our previous works~\cite{Kim:2017nxg} for photoproductions as follows: 
\begin{equation}
\label{eq:SFF}
F_x=\frac{\Lambda^4}{\Lambda^4+(x-M^2_x)^2}\,\,\,\,\mathrm{for}\,\,\,\,x=s,t,u.
\end{equation}
Here, $x$ indicates the Mandelstam variables. The strong cutoff parameter $\Lambda$ will be determined to reproduce the data in the next Section. Since a naive usage of form factors in Eq.~(\ref{eq:INVTOT}) breaks the WTI, we follow the prescription developed in Refs.~\cite{Haberzettl:1997jg,Haberzettl:1998eq,Davidson:2001rk}. A real photon-beam induced scattering amplitude, which consists of $s$, $t$, and $u$ channels, can be written in principle as follows:
\begin{equation}
\label{eq:AMPBARE}
i\mathcal{M}^\mathrm{bare}_\mathrm{total}=(i\mathcal{M}_{Es}+i\mathcal{M}_{Ms})
+(i\mathcal{M}_{Et}+i\mathcal{M}_{Mt})+(i\mathcal{M}_{Eu}+i\mathcal{M}_{Mu}),
\end{equation}
where the subscripts $E$ and $M$ stand for the electric $(\propto \rlap{/}{\epsilon})$ and magnetic $(\propto \rlap{/}{\epsilon}\rlap{/}{k}_\gamma)$ photon couplings. Thus, the magnetic amplitudes are \textit{self-gauge invariant}. In Refs.~\cite{Kim:2017nxg}, this amplitude was reorganized with the form factors into the \textit{dressed} one to satisfy the WTI as follows:
\begin{equation}
\label{eq:DAMP}
i\mathcal{M}^\mathrm{dressed}_\mathrm{total}=(i\mathcal{M}_{Es}+i\mathcal{M}_{Et}+i\mathcal{M}_{Eu})F_c+
i\mathcal{M}_{Ms}F_s+i\mathcal{M}_{Mt}F_t+i\mathcal{M}_{Mu}F_u.
\end{equation}
Note that $F_c$ denotes a \textit{common} form factor to save the identity as given by
\begin{equation}
\label{eq:CFF}
F_c=1-(1-F_s)(1-F_t),
\end{equation}
which satisfies the on-shell condition and the crossing symmetry. Since the sum of the electric contributions in Eq.~(\ref{eq:DAMP}) and each magnetic contribution becomes zero for the replacement $\epsilon_\gamma\to k_\gamma$ in the scattering amplitude, the WDI is fulfilled in this prescription.  

Being different from photoproductions as studied in Refs.~\cite{Kim:2017nxg}, electroproductions need photon-virtuality ($Q^2$) dependent form factors, i.e., EM form factors for hadrons. We have three EM vertices in the relevant Feynman diagrams in Fig.~\ref{FIG0} where the photon couples to the proton, the kaon, and the $\Lambda^*$. Before going further, we want to revisit the nucleon EM form factors briefly. Conventionally, there have been two interrelated descriptions: The Sachs $(G_{E,M})$, and the Dirac $(F_1)$ and Pauli $(F_2)$ form factors. By using the Sachs ones, the interference between the electric and magnetic contributions disappears for instance. They are interrelated via the Rosenbluth formula as follows:
\begin{equation}
\label{eq:GEGM}
G_E(Q^2)=F_1(Q^2)-\kappa_N\tau F_2(Q^2),
\,\,\,\,G_M(Q^2)=\mu_NG_E(Q^2)=F_1(Q^2)+\kappa_NF_2(Q^2),
\end{equation}
where $\tau=Q^2/4M^2_p$ and $\kappa_N$ stands for the anomalous magnetic moment $[\mu_N]$. Those form factors are normalized at $Q^2=0$ for the proton and neutron by
\begin{eqnarray}
\label{eq:GEMPN}
G^p_E(0)&=&1,\,\,\,\,G^p_M(0)=\mu_p,\,\,\,\,F^p_{1}(0)=1,\,\,\,\,F^p_2(0)=1,\,\,\,\,\mu_p-1=\kappa_p,
\cr
G^n_E(0)&=&0,\,\,\,\,G^n_M(0)=\mu_n,\,\,\,\,F^n_{1}(0)=0,\,\,\,\,F^n_2(0)=1,\,\,\,\,\mu_n=\kappa_n.
\end{eqnarray}
Straightforwardly, the Dirac and Pauli form factors are redefined with the Sachs ones for our purpose as follows:
\begin{equation}
\label{eq:EMFFPA}
F_1(Q^2)=\frac{G_E(Q^2)+\tau G_M(Q^2)}{(1+\tau)},\,\,\,\,
F_2(Q^2)=\frac{G_M(Q^2)-G_E(Q^2)}{\kappa_N(1+\tau)}.
\end{equation}
Note that, in many literatures, the Sachs form factors have been parameterized for the proton and the neutron by
\begin{equation}
\label{eq:EMFFPARA}
G^p_E(Q^2)\simeq G_D(Q^2),
\,\,\,\,
G^p_M(Q^2)\simeq \mu_pG_D(Q^2),
\,\,\,\,
G^n_E(Q^2)\simeq -\frac{a\mu_n\tau}{1+b\tau}G_D(Q^2),
\,\,\,\,
G^n_M(Q^2)\simeq \mu_nG_D(Q^2)
\end{equation}
with the dipole-type form factor
\begin{equation}
\label{eq:DFF}
G_D(Q^2)=\left[\frac{1}{1+Q^2\langle r^2\rangle^p_E/12}\right]^2,
\end{equation}
where $\langle r^2\rangle_{Ep}$ denotes the electric root-mean-squared (rms) radius for the proton, which is about $(0.863\pm0.004)\,\mathrm{fm}$~\cite{Kelly:2004hm} and provides good agreement with experimental data with the above parameterization~\cite{Kelly:2004hm}. In this work, we employ $F^p_{1,2}$ for the proton EM form factors in Eq.~(\ref{eq:EMFFPARADP}) for the numerical calculations. The most peculiar one is the neutron electric form factor $G^n_E$, which was usually given in terms of the Galster parameterization as shown Eq.~(\ref{eq:EMFFPARA})~\cite{Galster:1971kv}. The additional parameters are given by $(a,b)=(0.89\pm0.02,3.30\pm0.32)$~\cite{Kelly:2004hm}. Note that, in Ref.~\cite{Kaskulov:2003bg}, in which the authos considered the chiral contents in the neutron, $G^n_E$ was parameterized as follows:
\begin{equation}
\label{eq:AAAAA}
G^n_E(Q^2)=-\frac{\langle r^2\rangle^n_E}{6}Q^2F_\pi(Q^2)G_D(Q^2).
\end{equation}
Note that there are no phenomenological free parameters in this parameterization in principle, because $F_\pi(Q^2)$ and $\langle r^2\rangle^n_E$ can be extracted from experiments.

As noticed above, there have been many experimental and theoretical researches for the neutron EM form factor and its phenomenological parameterizations, whereas the information for the $\Lambda(1405)$ EM form factor is scarce. Hence, as a first trial to investigate the $\Lambda(1405)$ electroproduction, considering the charge neutralness of the $\Lambda(1405)$ and the neutron as a common feature, we assumed that the $\Lambda(1405)$ EM form factor follows the same parameterization with that for the neutron, and the difference in their internal structures is represented by their distinctive EM root-mean-squared (rms) radii, which appear in the parameterization.

Therefore, taking into account that the EM rms radii for $(H,L)$ and the Sachs form factors satisfy the following relations in general
\begin{equation}
\label{eq:RMS}
\langle r^2\rangle^{H,L}_E=-6\frac{dG^{H,L}_E(Q^2)}{dQ^2}\Big|_{Q^2=0},\,\,\,\,
\langle r^2\rangle^{H,L}_M=-\frac{6}{\mu_{H,L}}
\frac{dG^{H,L}_M(Q^2)}{dQ^2}\Big|_{Q^2=0},
\end{equation}
we may parameterize the EM form factors for $\Lambda^*=(H,L)$ as follows~\cite{Kaskulov:2003bg}:
\begin{equation}
\label{eq:EMHL}
G^{H,L}_E(Q^2)=-\frac{\langle r^2\rangle^{H,L}_E}{6}Q^2
F_K(Q^2)\left[\frac{1}{1+Q^2\langle r^2\rangle^{H,L}_M/12}\right]^2,\,\,\,\,
G^{H,L}_M(Q^2)\approx\mu_{H,L}
\left[\frac{1}{1+Q^2\langle r^2\rangle^{H,L}_M/12}\right]^2,
\end{equation}
which are normalized as $G^{H,L}_E(0)=0$ and $G^{H,L}_M(0)=\mu_{H,L}$. Note that we did not use the Galster parameterization in Eq.~(\ref{eq:EMFFPARA}) for $G^{H,L}_E(Q^2)$, since the parameterization suggested in Ref.~\cite{Kaskulov:2003bg} contains less parameters in principle. The values of the EM rms radii for the $H$ and $L$ are given in Table~\ref{TAB1} from the ChUM calculations~\cite{Sekihara:2008qk}. For comparison, we list the electric rms radius for the neutron as well. By comparing those values in Table~\ref{TAB1}, we expect that the high-mass pole $H$ is softer than the neutron, and vice versa for the low-mass one $L$. It is worth mentioning that the values of the EM rms radii in  Table~\ref{TAB1} were obtained at $M_{H,L}=(1426.11-16.65i,1390.43-66.21i)$, and these pole positions are rather inconsistent with those we employed in the present work as shown in Table~\ref{TAB0}. Therefore, from a phenomenological point of view, we assumed that the inconsistency can be compensated by other ingredients, such as the strong for factors, etc, after fitting the numerical results with the data. 
\begin{table}[b]
\begin{tabular}{|c|c|c|c|c|}
\hline
$\langle r^2\rangle^H_E$&
$\langle r^2\rangle^H_M$&
$\langle r^2\rangle^L_E$&
$\langle r^2\rangle^L_M$&
$\langle r^2\rangle^n_E$\\
\hline
$-3.365+7.783i$&
$6.859-10.455i$&
$0.462-0.051i$&
$-0.334+0.539i$&
$-2.877\pm0.077$\\
\hline
\end{tabular}
\caption{Electric ($E$) and magnetic ($M$) charge rms radii [1/GeV$^2$] for $(H,L)$ from the ChUM results~\cite{Sekihara:2008qk}. For comparison, we also show the neutron electric rms radius.}
\label{TAB1}
\end{table}
The EM form factor for the charged kaon $K^+$ reads~\cite{Jaus:1991cy}
\begin{equation}
\label{eq:KFFPARA}
F_K(Q^2)=\frac{1}{[1+Q^2/(0.845\,\mathrm{GeV})^2+Q^4/(1.270\,\mathrm{GeV})^4]},
\end{equation}
which gives $\langle r^2\rangle^{K^+}_E=0.327\,\mathrm{fm}^2$. By using those parameterized Sachs form factors for $(H,L)$, we define the Dirac and Pauli form factors, being similar to the nucleons, as follows:
\begin{equation}
\label{eq:EMFFPARADP}
F^{H,L}_1(Q^2)=\alpha_{H,L}
\left[\frac{G^{H,L}_E(Q^2)+\tau\,G^{H,L}_M(Q^2)}{\left(1+\tau\right)}\right],\,\,\,\,
F^{H,L}_2(Q^2)=\frac{G^{H,L}_M(Q^2)-G^{H,L}_E(Q^2)}{\mu_{H,L}\left(1+\tau\right)},
\end{equation}
and they satisfy the normalization conditions $F^{H,L}_1(0)=0$ and $F^{H,L}_2(Q^2)=1$ by constructions. Note that we have introduced a free parameter $\alpha_{H,L}$, because it does not make any effect on the normalization conditions for $F^{H,L}_1$ at $Q^2=0$. In the present work, we choose $\alpha_{H,L}=4$ as a trial to obtain a meaningful numerical results. Further theoretical and experimental studies for its Dirac form factors will shed light on determining the value of $\alpha_{H,L}$. By construction, the effects of the Dirac form factors become stronger as $\alpha_{H,L}$ increases.

\begin{figure}[t]
\includegraphics[width=8.5cm]{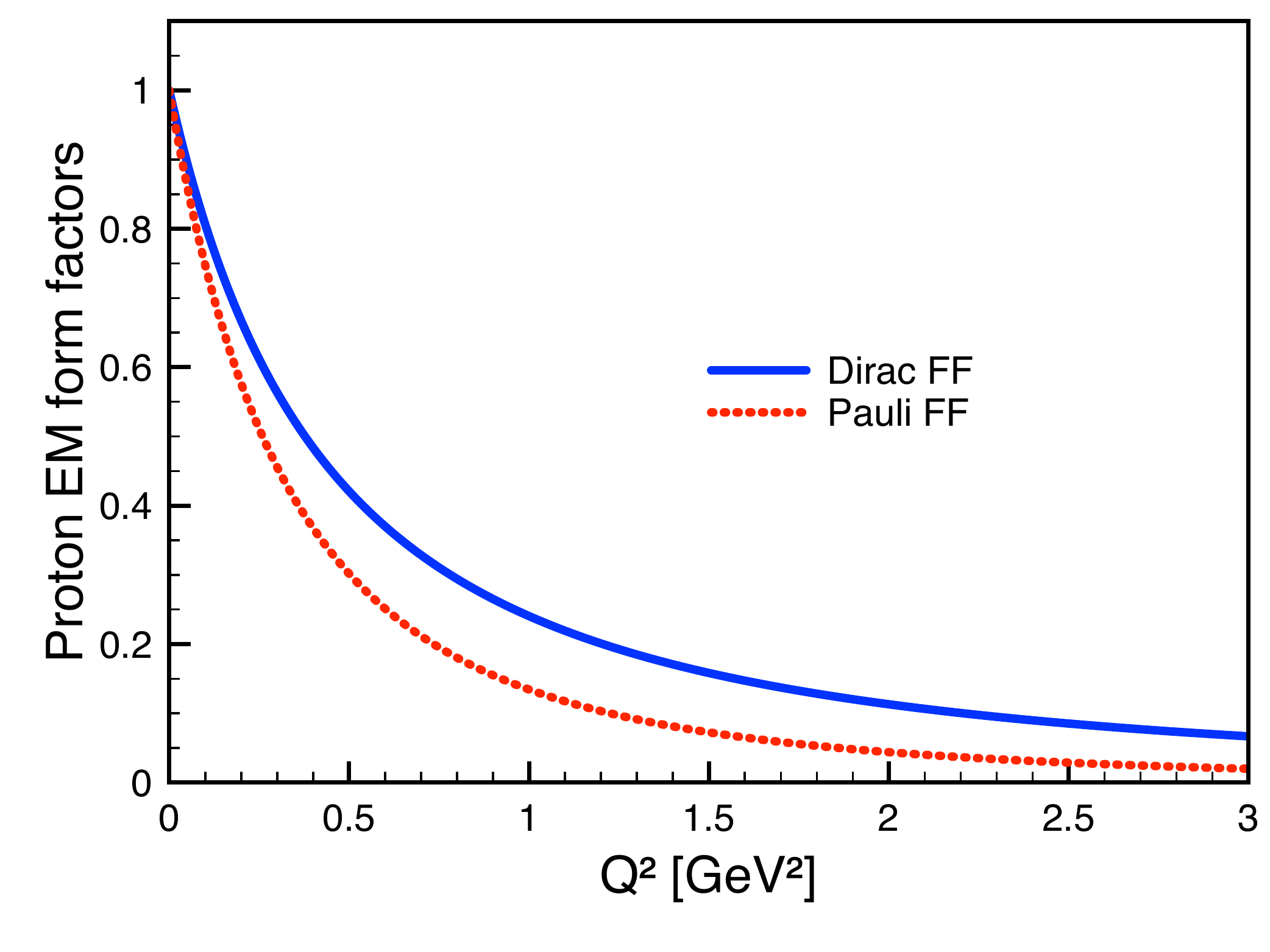}
\includegraphics[width=8.5cm]{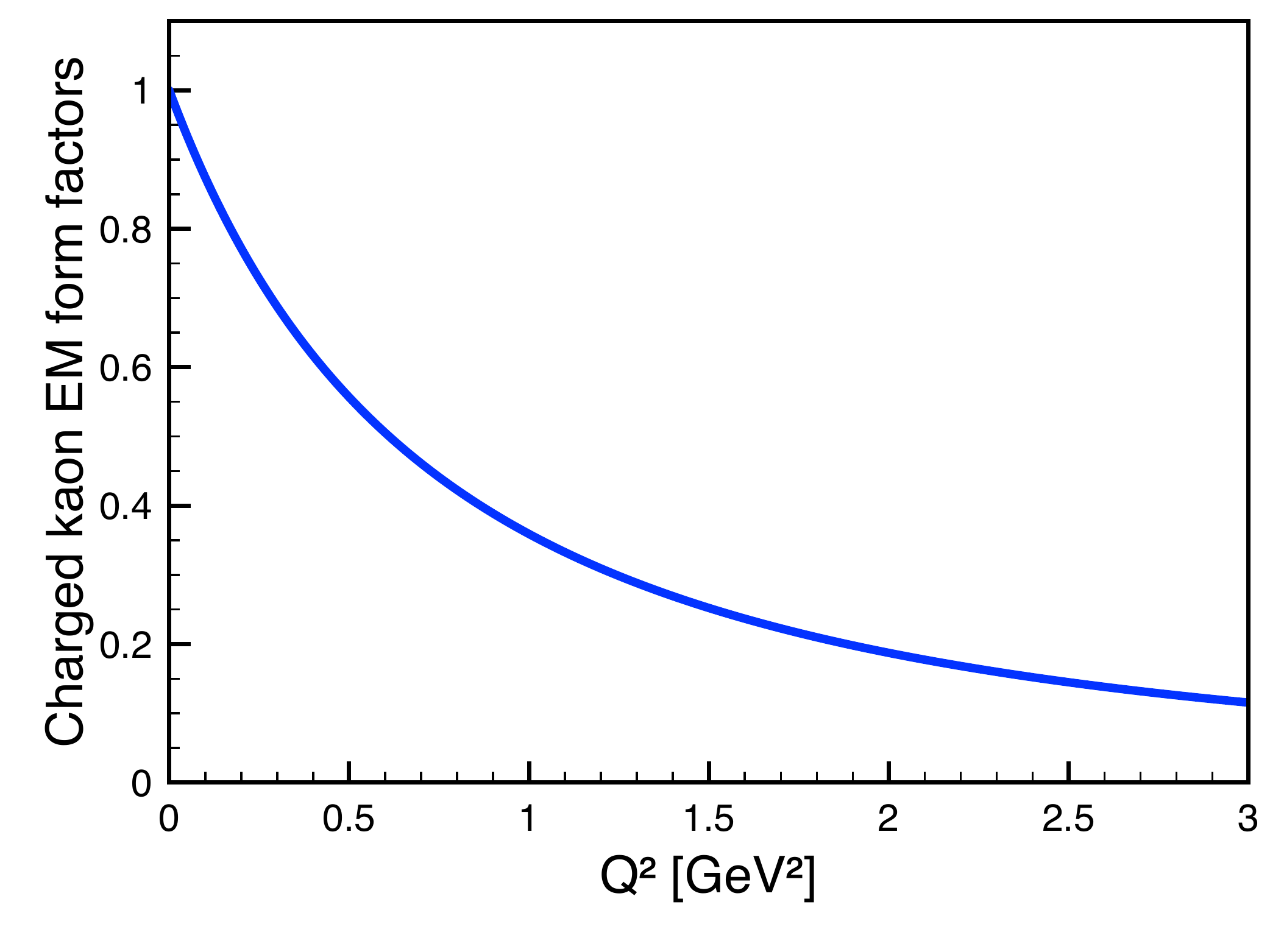}
\caption{(Color online) Left: Dirac (solid) and Pauli (dashed) form factors (FF) for the proton as functions of $Q^2$. Right: Electric form factor for the charged kaon as a function of $Q^2$. See the text for details.}       
\label{FIGPK}
\vspace{0.5cm}
\begin{tabular}{cc}
\includegraphics[width=8.5cm]{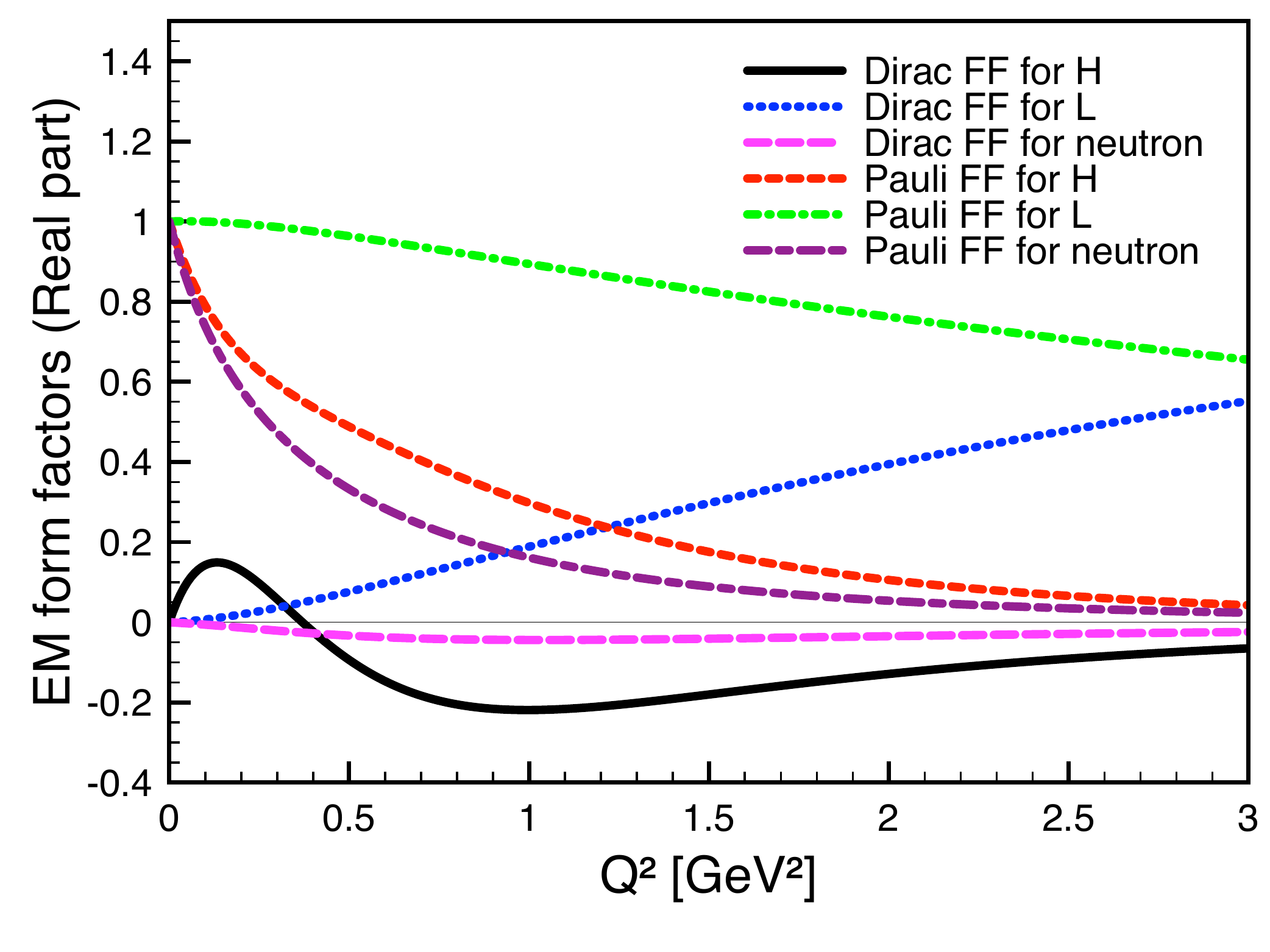}
\includegraphics[width=8.5cm]{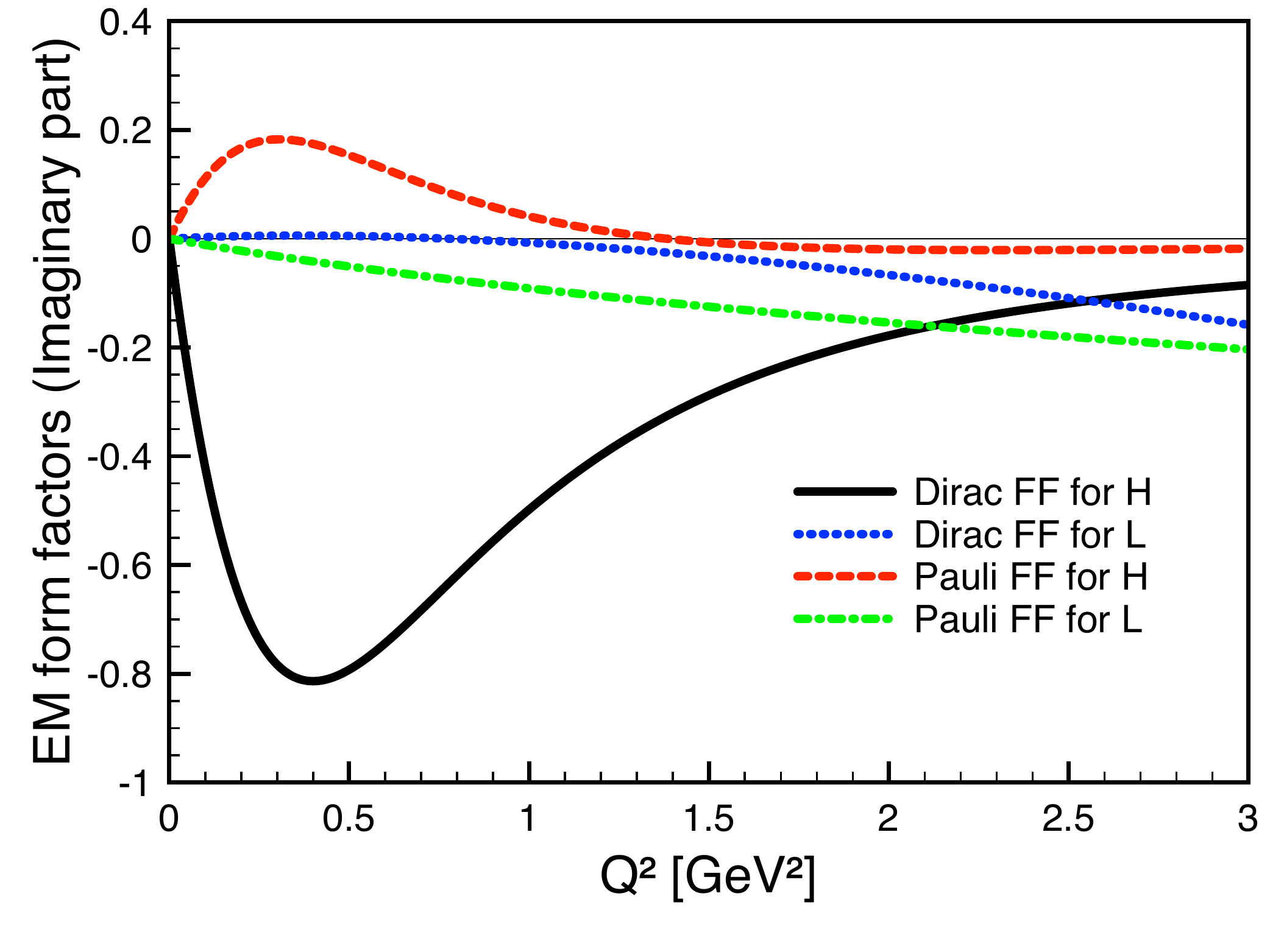}
\end{tabular}
\caption{(Color online) Left: The real parts of the Dirac $(F_1)$ and Pauli $(F_2)$ form factors (FF) for the high-mass ($H$) and low-mass ($L$) poles. We also show those form factors for the neutron. Right: The imaginary parts for the form factors.}       
\label{FIG3}
\end{figure}

In Fig.~\ref{FIGPK}, we show the numerical results for the proton (left) and charged kaon (right) EM form factors as functions of $Q^2$, satisfying the normalization conditions. The numerical curves for the Dirac $(F^{H,L}_1)$ and Pauli $(F^{H,L}_2)$ form factors for the real (left) and imaginary (right) parts in Fig.~\ref{FIG3}. In the left panel of Fig.~\ref{FIG3}, we also draw the neutron Dirac (long-dashed) and Pauli (long-dot-dashed) form factors for comparison. We observe that the EM form factors for $H$ behave relatively similar to those for the neutron, whereas those for $L$ are very different from the neutron EM form factors. This observation indicate the distinctive EM internal structures of $\Lambda^*$ in comparison to usual baryons. Moreover, the signs of the Dirac form factors for $H$ and $L$ starts to be opposite beyond $Q^2\gtrsim0.5\,\mathrm{GeV}^2$, while those for the Pauli ones remain positive. We will show that this sign oppositeness between $F^{H,L}_1$ makes changes in the interference patterns between the two poles in the electroproduction. 
\section{Numerical results and Discussions}
\begin{figure}[t]
\begin{tabular}{cc}
\includegraphics[width=8.5cm]{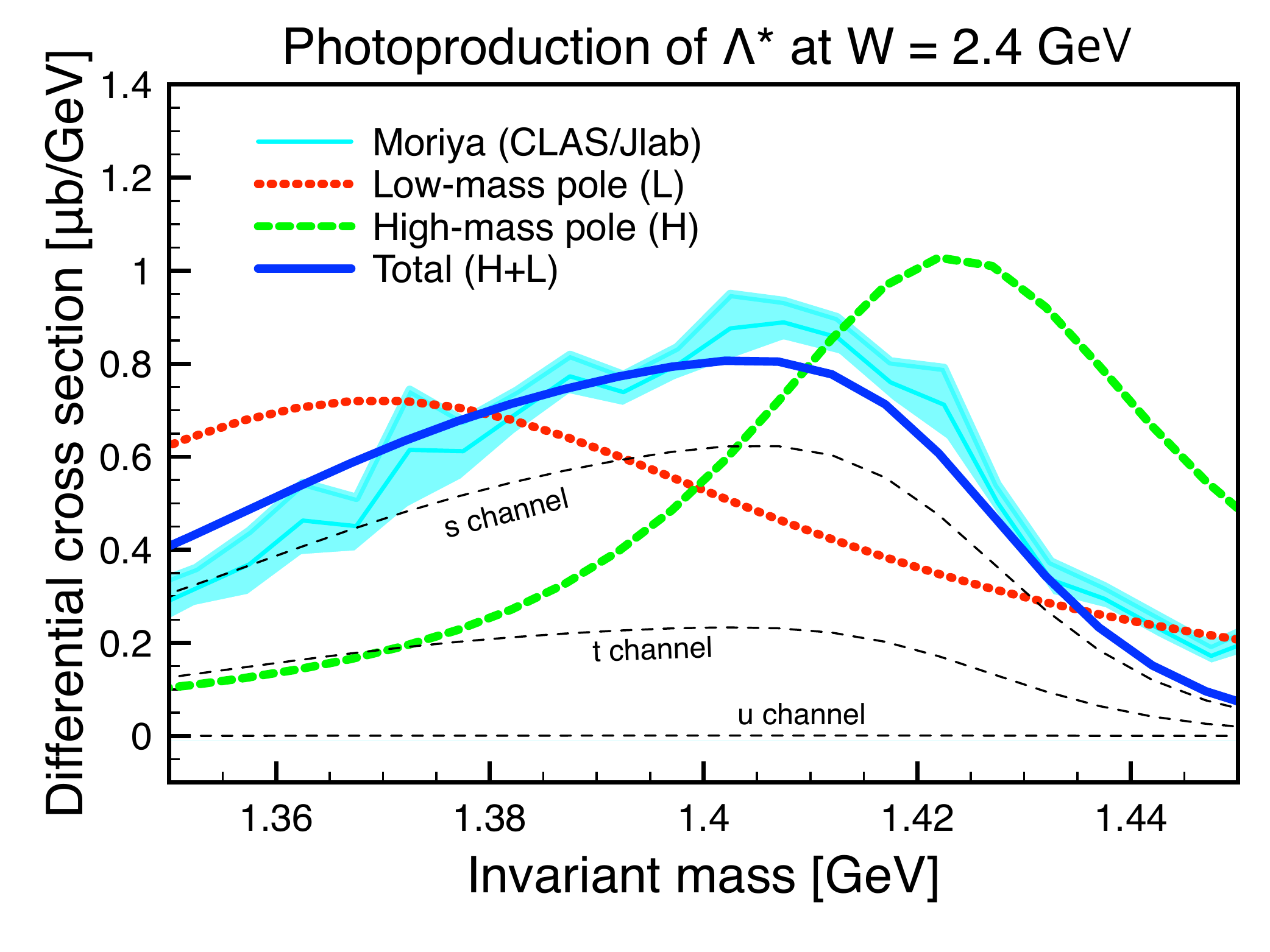}
\includegraphics[width=8.5cm]{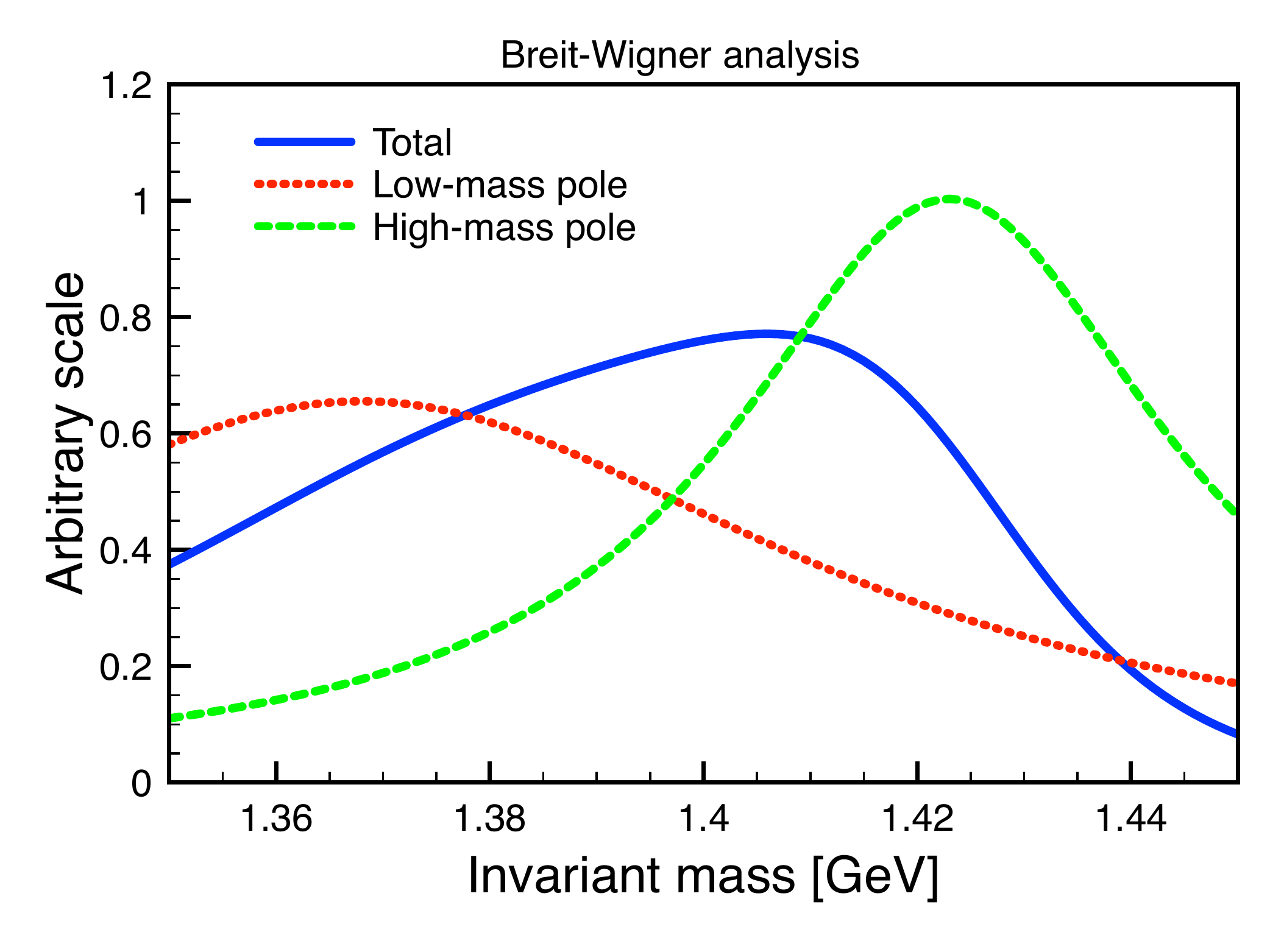}
\end{tabular}
\caption{(Color online) Left: Differential cross section $d\sigma_{\gamma p\to K^+\pi^+\Sigma^-}/dM_{\pi^+\Sigma^-}$ [$\mu\mathrm{b/GeV}$] as a function of the invariant mass $M_{\pi^+\Sigma^-}$ for the $\Lambda^*$ \textit{photoproduction} at $W=2.4\,\mathrm{GeV}$ for the contributions from $H+L$ (thick solid), $H$ (thick dashed), and $L$ (thick dotted). The photoproduction-experiment data are taken from CLAS/Jlab~\cite{Moriya:2013hwg}, where the shaded area indicate the errors. We also show the contributions from the $s$, $t$, and $u$ channels separately in the thin dashed lines. Right: The Breit-Wigner distribution analysis using Eq.~(\ref{eq:BW}) shows a destructive interference between the two poles.}       
\label{FIG1}
\end{figure}

In this Section, we provide numerical results for the photo- and electroproduction in the present toy-model and relevant discussions. First, we show the numerical results for the differential cross section for the photoproduction, i.e., $d\sigma_{\gamma p\to K^+\pi^+\Sigma^-}/dM_{\pi^+\Sigma^-}$ in the left panel of Fig.~\ref{FIG1}. As mentioned previously, to avoid the nucleon-resonance contributions, we perform the calculation at $\sqrt{s}\equiv W=2.4$ GeV, which is beyond the resonance region. The experimental data are taken from the CLAS/Jlab photoproduction data~\cite{Moriya:2013hwg}. Our full result is given by the solid line (total) and shows a good agreement with the data. For this, we have chosen the cutoff parameter for the strong form factors in Eq.~(\ref{eq:SFF}) to be $900$ MeV, which is similar to that determined for the the two-body reaction process $\gamma p\to K^+\Lambda^*$ in our previous work~\cite{Kim:2017nxg}. We also present the separate contributions from the high- and low-mass poles in the dashed and dotted lines, respectively, in addition to the each channel contributions in thin dashed lines. Note that the $s$- and $t$-channel contributions give the typical single-peak structure, whereas the $u$ channel one is almost negligible. By comparing the total and the separated $H$ and $L$ curves, we can conclude that the total curve shape is resulted from the destructive interference between the two poles, resulting in the single-pole line shape. As mentioned, the main reason for this interference can be understood by the overall \textit{sign-oppositeness} of the combined coupling constants as follows:
\begin{equation}
\label{eq:COMSTRONG}
g_{KNL}\,g_{\pi\Sigma L}=-0.450-6.050i,\,\,\,\,g_{KNH}\,g_{\pi\Sigma H}=0.266+3.895i
\end{equation}
This destructive interference can be easily tested by the simple Breit-Wigner type distribution analysis:
\begin{equation}
\label{eq:BW}
i\mathcal{M}_H=\frac{g_{KNH}\,g_{\pi\Sigma H}A_H}{M^2_{\pi^+\Sigma^-}-M^2_H-i\Gamma_HM_H},\,\,\,\,
i\mathcal{M}_L=\frac{g_{KNL}\,g_{\pi\Sigma L}A_L}{M^2_{\pi^+\Sigma^-}-M^2_L-i\Gamma_LM_L}.
\end{equation}

In the right panel of Fig.~\ref{FIG1}, we plot $|\mathcal{M}_H+\mathcal{M}_L|^2$ using Eqs.~(\ref{eq:COMSTRONG}) and (\ref{eq:BW}) for $A_H=A_L$ with a certain normalization. From the curves, we can clearly see the destructive interference between the high- and low-mass poles cause the typical single-pole line shape for the $\Lambda^*$ invariant-mass distribution for its photoproduction. 

Now, we are in a position to consider the electroproduction case. As observed in the CLAS/Jlab electroproduction experiment~\cite{Lu:2013nza}, there appear two peaks in the invariant-mass distribution for $Q^2=(1\sim3)\,\mathrm{GeV}^2$. Note that the proton and kaon EM form factors do not make any difference between the high- and low-mass poles, because those form factors have the same effects commonly on the vertices with $H$ and $L$. Therefore, the two peaks in the electroproduction should be the consequence of the two-pole EM form factors, which appear only in the $u$-channel contributions. From our parameterization of those Dirac form factors at $Q^2=2\,\mathrm{GeV}^2$ in Eq.~(\ref{eq:EMFFPARADP}) with the combined strong couplings in Eq.~(\ref{eq:COMSTRONG}), we find the EM-modified couplings in the $u$ channel as follows:
\begin{equation}
\label{eq:COMSTRONGDIRAC}
g_{KNL}\,g_{\pi\Sigma L}\,F^L_{1}=0.660-0.549i,\,\,\,\,
g_{KNH}\,g_{\pi\Sigma H}\,F^H_{1}=-0.583-2.363i.
\end{equation}
Note that the signs for the combined couplings are changed for the real parts due to the Dirac form factors, whereas the imaginary parts have the same negative signs with Eq.~(\ref{eq:COMSTRONG}). Hence, this observation indicates that the relative phase between the two poles is altered in comparison to that for the photoproduction, resulting in a different interference pattern. On the contrary, the Pauli form factors in the $u$ channel do not make any difference in the signs at $Q^2=2\,\mathrm{GeV}^2$, comparing to the photoproduction as follows:
\begin{equation}
\label{eq:COMSTRONGPAULI}
g_{KNL}\,g_{\pi\Sigma L}\,F^L_2=-1.275-4.536i,\,\,\,\,
g_{KNH}\,g_{\pi\Sigma H}\,F^H_2=0.105+0.405i.
\end{equation}
Thus, we can expect negligible differences in the interference pattern due to the Pauli form factors, although the strengths change. These observations are summarized in Table~\ref{TAB2}.
\begin{table}[b]
\begin{tabular}{|c|c||c|c||c|c|}
\hline
\multicolumn{2}{|c||}{Photoproduction}&
\multicolumn{2}{c||}{Electroproduction; Dirac }&
\multicolumn{2}{c|}{Electroproduction: Pauli}
\\
\hline
$g_{KNL}\,g_{\pi\Sigma L}$&
$g_{KNH}\,g_{\pi\Sigma H}$&
$g_{KNL}\,g_{\pi\Sigma L}F^L_{1}$&
$g_{KNH}\,g_{\pi\Sigma H}F^H_{1}$&
$g_{KNL}\,g_{\pi\Sigma L}F^L_2$&
$g_{KNH}\,g_{\pi\Sigma H}F^H_2$\\
\hline
$-0.450-6.050i$&
$0.266+3.895i$&
$0.660-0.549i$&
$-0.583-2.363i$&
$-1.275-4.536i,$&
$0.105+0.405i$\\
\hline
\end{tabular}
\caption{Combined and EM-modified strong coupling constants for the two poles for the photo- and electroproduction. Here, we choose $Q^2=2\,\mathrm{GeV^2}$ for $F_{1,2}$.}
\label{TAB2}
\end{table}

\begin{figure}[t]
\begin{tabular}{cc}
\includegraphics[width=8.5cm]{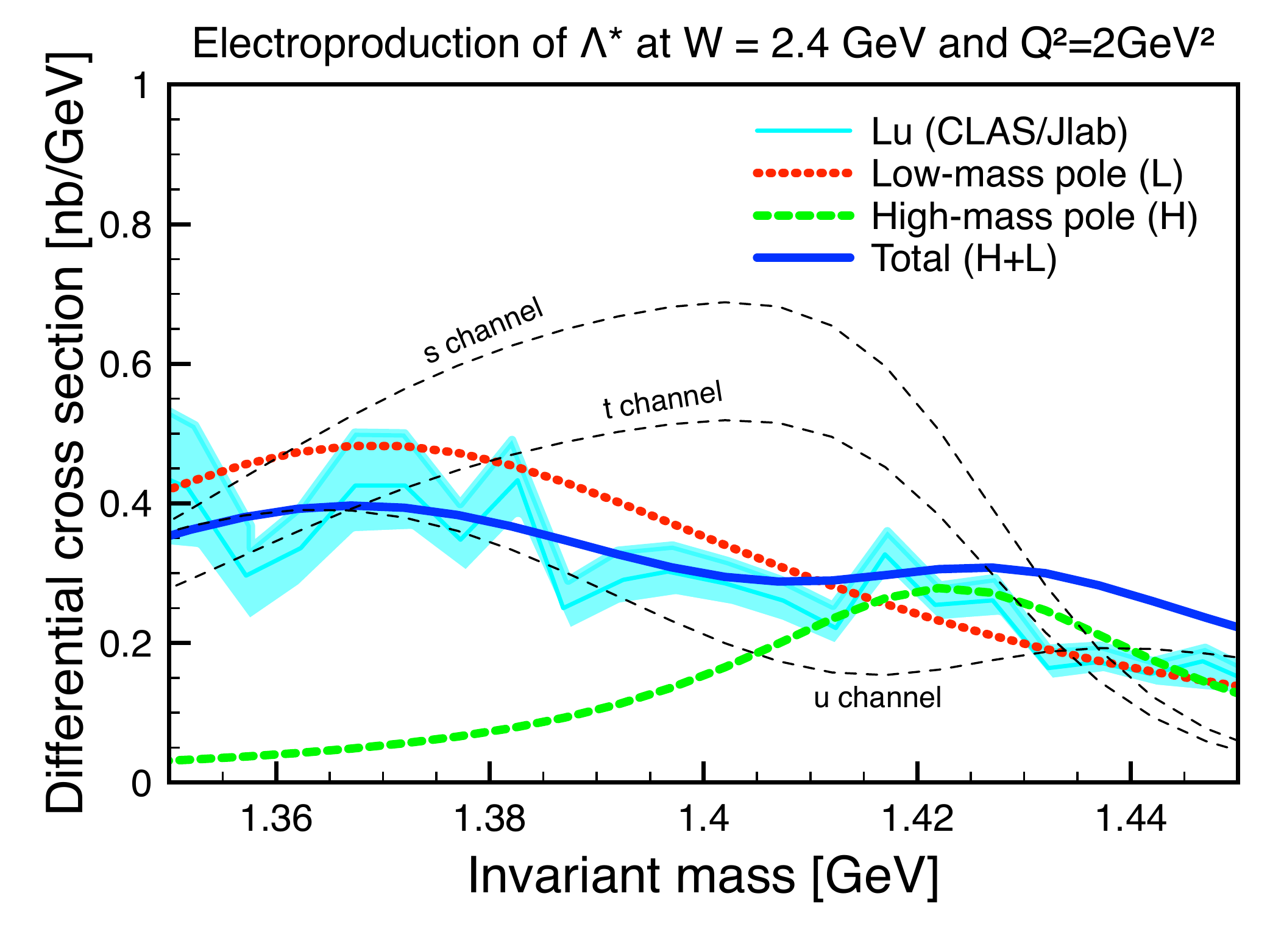}
\includegraphics[width=8.5cm]{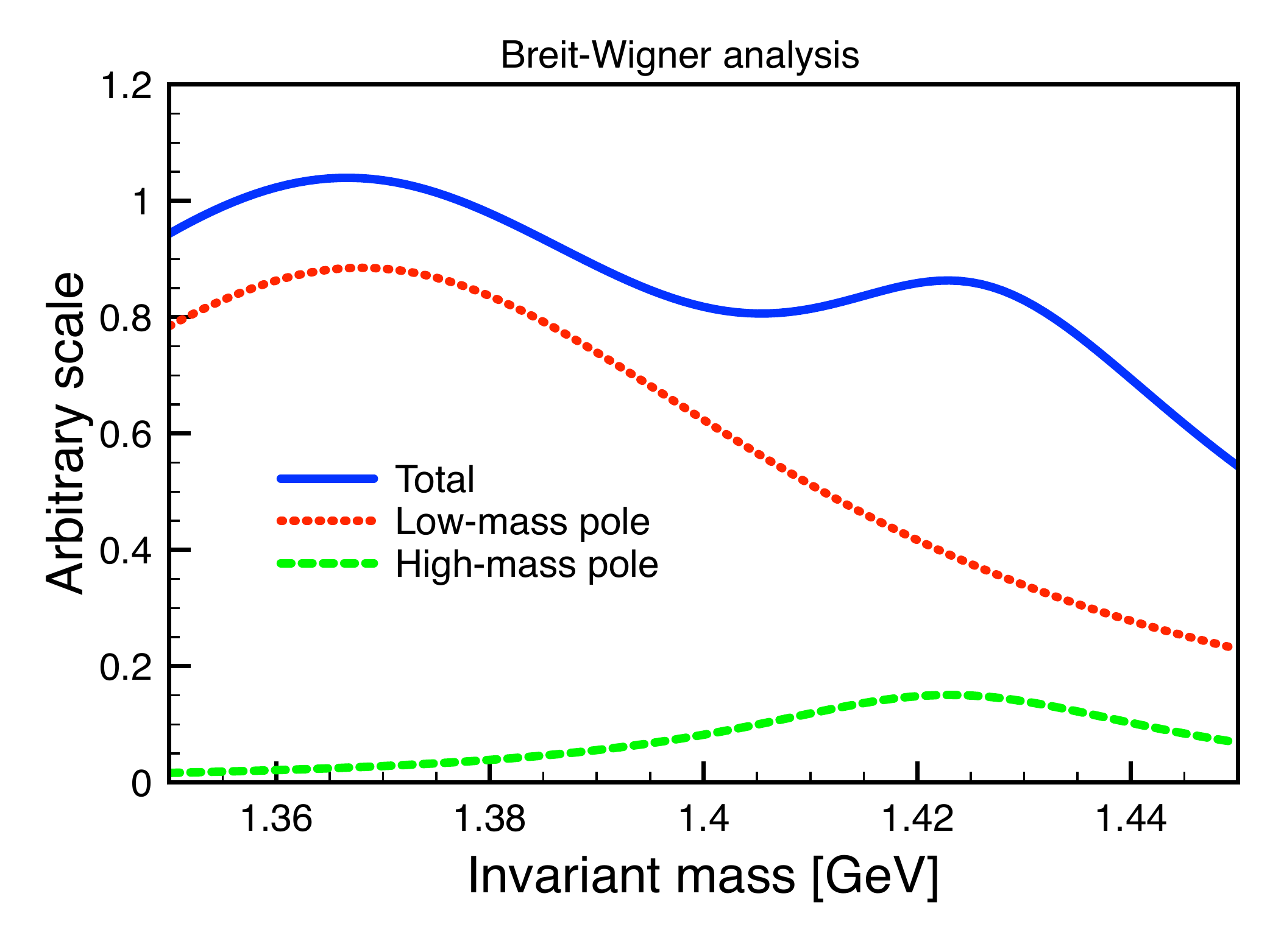}
\end{tabular}
\caption{(Color online) Left: Differential cross section $d\sigma_{\gamma^* p\to K^+\pi^+\Sigma^-}/dM_{\pi^+\Sigma^-}$ [$\mu\mathrm{b/GeV}$] as a function of the invariant mass $M_{\pi^+\Sigma^-}$ for the $\Lambda^*$ \textit{electroproduction} at $W=2.4\,\mathrm{GeV}$ for the contributions from $H+L$ (thick solid), $H$ (thick dashed), and $L$ (thick dotted). Here, we used $\varepsilon=0.5$ and $Q^2=2\,\mathrm{GeV}^2$. The experimental data for $1.0\,\mathrm{GeV}^2\le Q^2\le 3.0\,\mathrm{GeV}^2$ are taken from CLAS/Jlab~\cite{Lu:2013nza}. We also show the contributions from the $s$, $t$, and $u$ channels separately in the thin dashed lines. Right: The Breit-Wigner distribution analysis using Eq.~(\ref{eq:BW}) shows a destructive interference between the two poles.}       
\label{FIG2}
\end{figure}

In the left panel of Fig.~\ref{FIG2}, we show the numerical results for the electroproduction. We can clearly obtain the double-peak line shape (solid) in the invariant-mass distribution as observed in the CLAS/Jlab experiment~\cite{Lu:2013nza}. It is verified that the double-peak line shape depends much on the strength of the Dirac form factors, which appear only in the $u$ channel and change the interference pattern as discussed above, while the contributions with the Pauli form factors make the line shape similar to that for the photoproduction, i.e., the single-peak line shape as shown in Fig.~\ref{FIG1}. We also present the separate contributions from the two poles in the dotted ($L$) and the dashed ($H$) lines, in addition to each channel contributions in the thin dashed lines. As expected, the $s$- and $t$-channel contributions show the single-peak structure being similar to those for the photoproduction. On contrary, the $u$-channel one that contains the distinctive $H$ and $L$ EM form factors represents the double-peak structure describing the experiment qualitatively as we discussed above. At the same time, it can be understood that the magnetic contributions $\propto F_2$ are much smaller than the electric ones $\propto F_1$, taking into account Eq.~(\ref{eq:COMSTRONGPAULI}).

Thus, the constructive interference between the two poles provide the double-peak line shape in the electroproduction of $\Lambda^*$. Just for a qualitative comparison and ignoring the decay $K^*\to\pi\Sigma$, the experimental data for the $\Lambda(1405)$ electroproduction via $\gamma^*p\to K^+\pi^-\Sigma^+$~\cite{Lu:2013nza} are also given in the figure.  Being similar to the photoproduction, we also perform the BW-distribution analysis of $|\mathcal{M}_H-\mathcal{M}_L|^2$ using Eqs.~(\ref{eq:BW}) and (\ref{eq:COMSTRONGDIRAC}) with $A_H=A_L/3$  with a certain normalization and show the result in the right panel of Fig~\ref{FIG2}. Although the curve strengths are different from the full calculations in the left panel, we find that the constructive interference between the two poles give the double-peak line shape.

\begin{figure}[t]
\includegraphics[width=8.5cm]{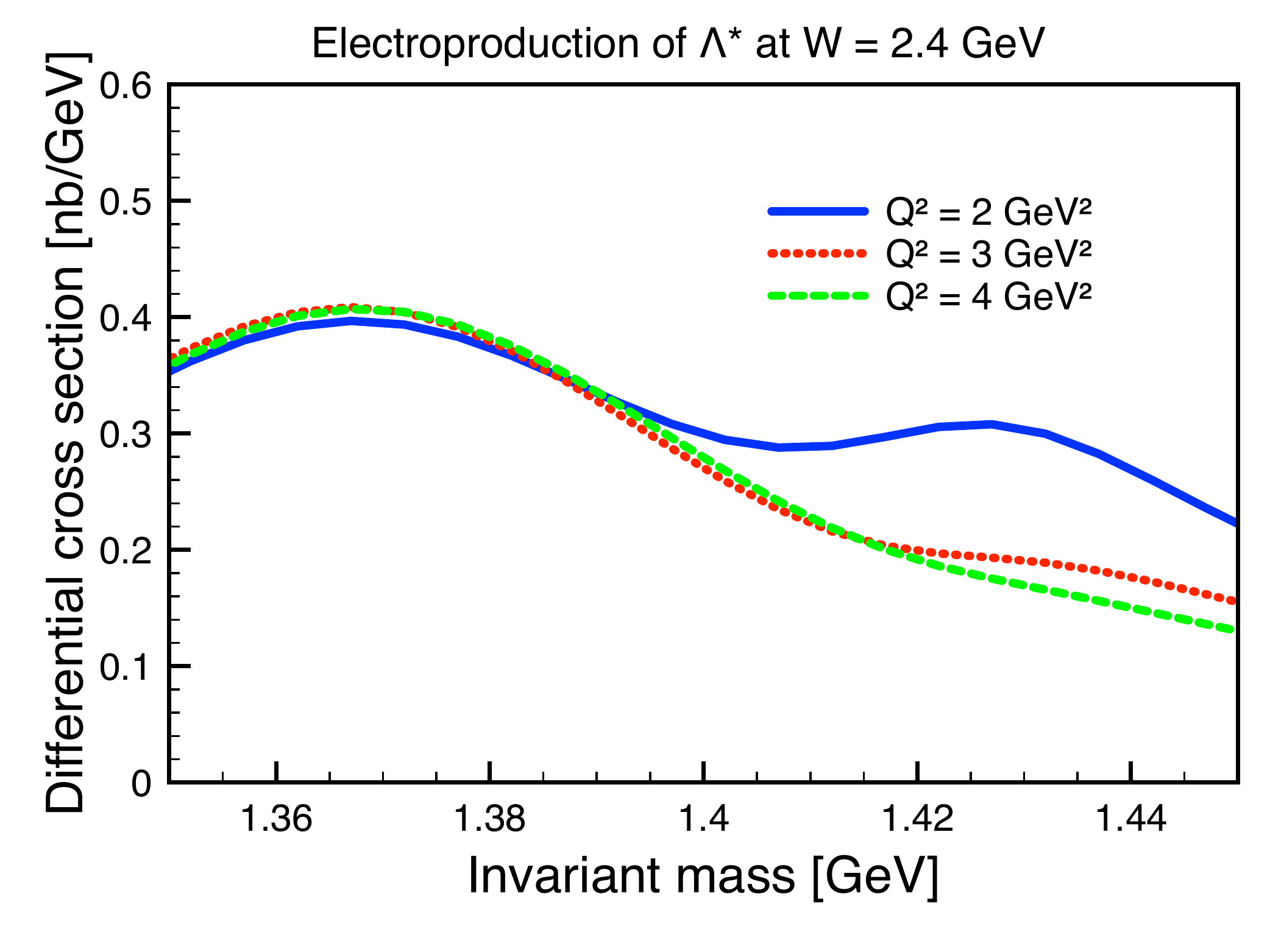}
\caption{(Color online) Differential cross section $d\sigma_{\gamma^* p\to K^+\pi^+\Sigma^-}/dM_{\pi^+\Sigma^-}$ [$\mu\mathrm{b/GeV}$] as a function of the invariant mass $M_{\pi^+\Sigma^-}$ for the $\Lambda^*$ \textit{electroproduction} at $W=2.4\,\mathrm{GeV}$ for $Q^2=2\,\mathrm{GeV}^2$ (solid), $3\,\mathrm{GeV}^2$ (dotted), and $4\,\mathrm{GeV}^2$ (dashed).}       
\label{FIG4}
\end{figure}

In Fig.~\ref{FIG4}, we show the same curves for $Q^2=(2\sim4)\,\mathrm{GeV}^2$ at $W=2.4$ GeV. In the electroproduction experiment of CLAS~\cite{Lu:2013nza}, the lower-pole contribution ($L$) becomes weaker in comparison to the higher one ($H$) as $Q^2$ increases. In contrast, our model result shows the weaker $H$ contribution with respect to $Q^2$. This observation indicates that the simple parameterizations for the EM form factors for $(H,L)$ in the previous Section still needs more realistic considerations. For instance, the factor $\alpha_{H,L}$ in Eq.~(\ref{eq:EMFFPARADP}) can be functions of $Q^2$: $\alpha_L(Q^2)$ decreases as $Q^2$ increases, and vice versa for $\alpha_H(Q^2)$. We would like to leave this interesting possibility for the future works, since we want to focus on the qualitative understanding for the two-peak structure shown in the electroproduction in the present work. 

Consequently, although the present model needs more realistic parameterizations for those EM form factors to reproduce the data correctly, we can conclude qualitatively that the distinctive EM interactions for the two poles make different line shapes for the photo- and electroproduction. Especially the electric couplings for the two poles, being represented by the Dirac form factors, play an important role to describe the line-shape difference, i.e. the single- and double-peak line shapes. It is worth mentioning that the present conclusion depends much on the theoretical estimates from the ChUM calculations. Because ChUM suggest the two poles and different EM properties for them, in terms of the consistency within ChUM, our findings are remarkable and support the scenario of the two-pole structure for $\Lambda^*$.

\section{Summary and future perspectives}
In the present work, we performed a toy-model analysis for the photo- and electroproduction of $\Lambda(1405)\equiv\Lambda^*$ in the Dalitz process, i.e., $\gamma^{(*)}p\to K^+\pi^+\Sigma^-$. We focussed on the experimental differences between the line shapes for the photo- and electroproduction, showing the single-peak and the double-peak line shapes, respectively. Thus, we tried to understand the reason for those obvious differences by considering the two-pole structure for $\Lambda^*$. For this purpose, we employed the effective Lagrangian method at the tree level. Considering these simplifications, we could clearly separate the $\Lambda^*$ contributions from others. All the model parameters were fixed by the theoretical and experimental information, especially from ChUM, which suggested the two-pole structure for $\Lambda^*$. Below, we list important achievements and observations in the present work:
\begin{itemize}
\item The analytical expression for the scattering amplitudes is constructed to satisfy the Ward-Takahashi identity naturally for the photoproduction as well as the electroproduction for the Dalitz process, i.e., $\gamma^{(*)}p\to K^+\pi^+\Sigma^-$. The strong and EM form factor are taken into account approproately. 
\item The photoproduction-experiment data from CLAS/Jlab are well reproduced within the model by modifying the full decay widths slightly from those by ChUM, and it shows that the typical tilted single-peak line shape can be understood by the destructive interference between the two poles, locating at $M_L=1368$ MeV and $M_H=1423$ MeV. 
\item Being based on the knowledge for the neutron form factor, we parameterize the Sachs form factors for the two poles by using the ChUM estimates for their electric and magnetic rms radii. Using the Sachs form factors, we define the Dirac and the Pauli form factors, which are inserted to our scattering amplitudes. We find that those EM form factors exhibits very different $Q^2$ dependences, comparing to that for the neutron, although they need more realistic $Q^2$ dependences to reproduce the data. 
\item We observe that the Dirac form factors for the two poles provide the different interference pattern for the electroproduction in comparison to that for the photoproduction, whereas the Pauli form factors do not make much differences. Because of this different pattern, i.e., the constructive interference between the two poles, the line shape becomes the double-peak one as reported by the CLAS/Jlab electroproduction experiment.
\end{itemize}

Consequently, the difference between the photo- and electroproduction line shapes is originated from the electric charge distributions of the two poles which are encoded by the Dirac form factors. In other words, the two poles exhibit different internal structures. In addition, because the difference can be explained well by the theoretical inputs estimated by ChUM, which suggested the two-pole structure for $\Lambda^*$, the difference in the line shapes must be a strong experimental and theoretical evidence for the two-pole exotic structure of $\Lambda^*$. Hence, experimental measurements for the EM form factors for $\Lambda^*$ will shed light on profound understandings for its genuine structure, although it must be very challenging. More realistic model calculations with nonresonant backgrounds, more hyperon resonances, various isospin channels for the $K^+\pi\Sigma$ are in progress and will appear elsewhere.
\section*{Acknowledgment}
S.i.N. thanks D.~Jido (TMU), S.~H.~Kim (APCTP), and T.~H.~Lee (PKNU) for fruitful discussions and helpful comments. The present work is partially supported by the research fund of Pukyong National University. The Feynman diagrams in this work were generated via \texttt{https://feynman.aivazis.com/}.

\end{document}